\documentclass[aps,eqsecnum,amsmath,onecolumn,notitlepage,nofootinbib]{revtex4-1}
\usepackage{graphics,setspace,epsfig,color}
\usepackage[letterpaper, dvips,width=7.5in,height=8.5in,includemp=false]{geometry}
\usepackage[vcentermath]{}
\usepackage{epsf}
\usepackage{amscd}
\usepackage{amsmath}
\usepackage{amsfonts}
\usepackage{color}

\usepackage[colorlinks=true,backref=false, linktocpage=true,
citecolor=blue,urlcolor=blue,linkcolor=blue,pdfpagemode=UseOutlines]{hyperref}

\newcommand{\beq}{\begin{equation}}
\newcommand{\eeq}{\end{equation}}
\newcommand{\bea}{\begin{eqnarray}}
\newcommand{\eea}{\end{eqnarray}}

\newcommand{\re}{\mathbb{R}}
\newcommand{\co}{\mathbb{C}}

\newcommand{\eq}[1]{Eq.~(\ref{#1})}
\newcommand{\fig}[1]{Fig.~\ref{#1}}
\newcommand{\tab}[1]{Table~\ref{#1}}

\def\tr {\mathop{\hbox{tr}}}

\begin{document}

\title[title]{Sign problem and Monte Carlo calculations beyond Lefschetz thimbles}
\author{Andrei Alexandru}
\email{aalexan@gwu.edu }
\affiliation{Department of Physics\\The George Washington University\\
Washington, DC 20052}
\author{G\"ok\c ce Ba\c sar}
\email{gbasar@umd.edu}
\author{Paulo F. Bedaque}
\email{bedaque@umd.edu}
\author{Gregory W. Ridgway}
\email{gridgwa1@umd.edu}
\author{Neill C. Warrington}
\email{ncwarrin@umd.edu}
\affiliation{Department of Physics \\
University of Maryland\\College Park, MD 20742}


\begin{abstract}
We point out that Monte Carlo simulations of theories with severe sign problems can be profitably performed over manifolds in complex space different from the one with fixed imaginary part of the action (``Lefschetz thimble"). We describe a family of such manifolds that interpolate between the tangent space at one critical point (where the sign problem is milder compared to the real plane but in some cases still severe) and the union of relevant thimbles (where the sign problem is mild but a multimodal distribution function complicates the Monte Carlo sampling). We exemplify this approach using a simple $0+1$ dimensional fermion model previously used on sign problem studies and show that it can solve the model for some parameter values where a solution using Lefschetz thimbles was elusive.
\end{abstract}

\maketitle

\section{Introduction}
The Monte Carlo method is frequently the only method applicable to strongly coupled field theoretical or many-body systems.
As  versatile as it is, a very important category of models has been up to now excluded from its reach, namely, problems where the action (or its euclidean version) is complex. Models in this category includes most field theory models (like QCD)  with a chemical potential, the repulsive Hubbard away from half-filling as well as real time observables. Roughly speaking, this problem arises from subtle cancellations among contributions with similar weight but differing phases (or signs) and it is termed ``the sign problem". Many ways to circumvent it have been proposed in the literature over the years with varied but limited amounts of success.

In  \cite{Cristoforetti:2012su} a new approach to the sign problem was suggested. It consists of complexifying all the variables of integration in the path integral and deforming the region of integration from the real values to a  manifold where the action is real (or has a constant imaginary part). The integrand over this other manifold, a union of the so-called ``Lefschetz thimbles",  has a fixed phase in each connected part.  In principle,  the integral over this manifold does not have a sign problem. Specific algorithms to perform the integration over a thimble were suggested and applied to a variety of simple toy models with bosonic degrees of freedom~\cite{Cristoforetti:2013wha,Fujii:2013sra,Mukherjee:2013aga,Cristoforetti:2013qaa,Cristoforetti:2014gsa,Fukushima:2015qza,Tanizaki:2014tua,Tanizaki:2015rda}. 

Fermionic models, where most of the interest in solving the sign problem lies, bring a new feature~\cite{Kanazawa:2014qma,Fujii:2015bua,Fujii:2015vha,Fujii:2015vha}. Some thimbles are bounded and end on a point where the fermion determinant vanishes and the effective action, which includes the $\tr\log(D)$ term, diverges. Other thimbles are connected to the same zero of the fermion determinant and they are, in turn, connected to further thimbles through their boundaries. It is not surprising then that the original integral over real fields is equivalent  to the integral over a large number of thimbles. This seems to be a nearly insurmountable difficulty for the application of the thimble idea to fermionic models. All the algorithms proposed for the computation of one thimble require the identification of the corresponding critical point but identifying a larger number of those in a realistic theory is a daunting process.  In addition, one would need to determine the combination of thimbles equivalent to the original real integral, a task of great complexity to say the least
 \footnote{As the action is an extensive quantity, arguments were made that only the thimble with the smallest action contributes in the thermodynamic limit. Also, universality has been used to claim that computing the contribution of one thimble is sufficient in the continuum limit. These observations are the basis of the philosophy advocated, for instance,  in \cite{Scorzato:2015qts}, stating that the integration over one thimble suffices in the continuum/thermodynamic limits.   The present paper can be seen as an alternative to this point of view as it suggests a practical way of performing the path integral even if not dominated by one thimble.}. 
The need for a multi-thimble integration, at least in some models, was evidenced  in~\cite{Alexandru:2015xva} where we considered a simple soluble $0+1$ dimensional  fermionic model previously used as a toy model for the sign problem~\cite{Pawlowski:2013pje,Aarts:2008rr}. We showed that an algorithm designed to compute the contribution of one thimble gave the exact answer as long as the coupling constant was small and the temperature and lattice spacing were not too small. However, at strong coupling or in the zero temperature or continuum limits, there were  small but statistically significant discrepancies from the exact result. Furthermore, a semiclassical estimate of the contribution from other thimbles matched, in order of magnitude, the size of the discrepancy. This result suggests  that the algorithm proposed in~\cite{Alexandru:2015xva} properly samples one thimble and that the discrepancies result from neglecting the other thimbles' contributions.

The objective of the present paper is to point out that the partition function can also be computed as an integration over other manifolds in complex space besides thimbles and that it is useful to do so. We consider, in particular, a  family of manifolds, parametrized by the flow time,  interpolating between thimbles and the tangent space of {\it one single} thimble. Each one of these manifolds is obtained by flowing the tangent space of {\it one} thimble.   We exemplify this observation using  the $0+1$ dimensional model mentioned above. We show that we can obtain the exact results both in the continuum limit and up to very low temperatures---something that eluded us in~\cite{Alexandru:2015xva} -- by simply integrating over the tangent space. In the case of the lowest temperatures, where the integration over the tangent space fails, the calculation is still possible if a different manifold, obtained by flowing by a moderate amount, is used.
 
%
%
%

\section{Thimbles and the Contraction Algorithm}
Expectation values of an observable $\mathcal{O}$ can be computed in field theory as a path integral of the form
\beq\label{eq:partition}
\langle  \mathcal{O}\rangle= \frac{1}{Z}\int D\phi\ e^{-S[\phi]} \mathcal{O}[\phi], \quad\text{with}\quad Z=\int D\phi\ e^{-S[\phi]},
\eeq 
where the integration is over real fields $\phi$. We assume from the start that some discretization is used so the integration is over $\re^N$, where $N$ is the number of degrees of freedom of the system (proportional to the number of spacetime points). The standard procedure for an stochastic calculation of $\langle  \mathcal{O}\rangle$ is to obtain a set of $\mathcal{N}$ field configurations $\{  \phi^{(a)}\}$ distributed according to the distribution $P[\phi] = e^{-S[\phi]}/Z$ and use them to estimate the expectation value of the observable as
\beq
\langle\mathcal{O}\rangle = \frac{1}{\mathcal{N}}  \sum_a \mathcal{O}[\phi^{(a)}].
\eeq 
Unfortunately, $e^{-S[\phi]}/Z$ cannot be interpreted as a probability distribution if $S$ is not real and stochastic  methods fail. A possible way around this problem is to use the real part of $S$ in the distribution probability and include the imaginary part in the operator to be averaged:
\beq
\langle  \mathcal{O}\rangle = 
 \frac{ \int D\phi\ e^{-S_R[\phi] }    e^{-i S_I[\phi]}  \mathcal{O}[\phi]  } 
        {   \int D\phi\ e^{-S_R[\phi]}    e^{-i S_I[\phi]} }
 = \frac{ \int D\phi\ e^{-S_R[\phi] }    e^{-i S_I[\phi]}  \mathcal{O}[\phi]  } 
        {   \int D\phi\ e^{-S_R[\phi]}   }
 \frac{ \int D\phi\ e^{-S_R[\phi] }    } 
        {   \int D\phi\ e^{-S_R[\phi]}    e^{-i S_I[\phi]} }       
 = \frac{ \langle   e^{-i S_I}  \mathcal{O}   \rangle_{S_R}}{\langle  e^{-i S_I}    \rangle_{S_R}}.
\eeq 
This ``reweighting" method works as long as the phase $e^{-i S_I}$ does not fluctuate too wildly from configuration to configuration. A measure of how much the phase fluctuates is the value of the average phase  $\langle  e^{-i S_I}    \rangle_{S_R}$. As it approaches zero a very large number of configurations are required for the computation of $\langle  \mathcal{O}\rangle$. Unfortunately, as $S$ is an extensive quantity, the fluctuation of the phase  is expected to be proportional to the spacetime volume (that is, it is proportional to the spatial volume and inversely proportional to the temperature) and one expects a wildly fluctuating phase in the low temperature and thermodynamic limits. This is the famous ``sign problem" that has impeded Monte Carlo calculations in theories of as great interest as QCD at finite chemical potential.
 
An idea proposed recently~\cite{Cristoforetti:2012su} is to complexify the variables of integration and change the region of integration away from $\re^N$ into a manifold of $N$ (real) dimensions immersed in $\co^N$ where the imaginary part of the action is constant  and the sign problem disappears. One such manifold is known as a ``Lefschetz thimble", that is, a multidimensional generalization of the ``stationary phase path" or the ``steepest descent direction" used in asymptotic expansions of one dimensional integrals. A Lefschetz thimble is defined for every critical point, that is, a field configuration $\phi_c$ (not necessarily real) satisfying the classical equations of motion (an account of these methods can be found in~\cite{fedoryuk,pham}; see also~\cite{Witten:2010zr}):
\beq
\left.\frac{\partial S}{\partial \phi_i}\right|_{\phi=\phi_c} \!\!=0.
\eeq 
The thimble of the critical point $\phi_c$ is defined as the set of points that, under the (downward) flow defined by the equations
\beq\label{eq:flow_eq}
\frac{d \phi_i}{dt} = - \overline{  \frac{\partial S}{\partial \phi_i} },
\eeq 
approach $\phi_c$ asymptotically. We can gain some insight into~\eq{eq:flow_eq} by splitting  $\phi_i = x_i + i y_i$ and $S=S_R + i S_I$  into real and imaginary parts:
\bea
\frac{d x_i}{dt} &=& - \frac{\partial S_R}{\partial x_i } = -\frac{\partial S_I}{\partial y_i },\\
\frac{d y_i}{dt} &=& - \frac{\partial S_R}{\partial y_i } = \frac{\partial S_I}{\partial x_i } .
\eea 
The first equality shows that the flow in~\eq{eq:flow_eq} is the (negative of the) gradient flow of the real part of $S$ and that the imaginary part of $S$ is constant along the flow (and, consequently, constant over each thimble). It is in this sense that thimbles are  generalizations of the path of steepest descent or stationary phase used in one-dimensional ($N=1$) integrals. 
 
The constancy of $S_I$ over a thimble is useful in addressing the sign problem as the $e^{-i S_I}$ factor can be factored out of the integral. In general, however, several thimbles need to be added together to reproduce the original integral over real fields. More precisely, the original integral over the real fields is given by
\beq\label{eq:thimble_decomposition}
\langle  \mathcal{O}[\phi]\rangle = 
\frac{  \sum_\sigma  n_\sigma e^{-i S_I[\phi_\sigma]}  \int_\sigma D\phi\ e^{-S_R[\phi]}   \mathcal{O}[\phi] }
 {  \sum_\sigma n_\sigma e^{-i S_I[\phi_\sigma]}  \int_\sigma D\phi\ e^{-S_R[\phi]}  },
\eeq 
where $\sigma$ indexes the different thimbles and $\phi_\sigma$ are the corresponding critical points. The integers $n_\sigma$ are determined by the crossing of the original integration region with the ``upward flow sets" defined as the set of points approaching a given critical point along the upward flow
\beq\label{eq:flow_up}
\frac{d \phi_i}{dt} = +\overline{ \frac{\partial S}{\partial \phi_i}  }.
\eeq 
The analysis of the flow equations~\eq{eq:flow_eq} near the critical point shows that thimbles are manifolds with $N$ real dimensions\footnote{We are assuming the non-degenerate case where the determinant of the matrix $\partial^2 S/\partial \phi_i\partial \phi_j$ at the critical point is not zero.}.
 
Thimbles in fermionic models have a distinct feature not present in bosonic models~\cite{Kanazawa:2014qma,Fujii:2015bua,Fujii:2015vha}. If one uses the standard procedure of introducing an auxiliary bosonic field through the Hubbard-Stratanovich transformation and then integrate over the fermion fields one is left with a  determinant $\det D[\phi]$ which, when exponentiated, adds to the auxiliary field action a term $-\log {\rm det} D[\phi]$. At the values of $\phi$ at which the determinant vanishes the  action has a logarithmic singularity and the flow equations cannot be integrated past a certain value of $t$. Thus, thimbles have boundaries at finite distances from the critical point\footnote{The submanifold of $\co^N$ with $\det D=0$ has real dimension $2N-2$ while the thimble has dimension $N$. Generically, in a $2N$ dimensional real manifold ($\co^N$), their intersection is an $N-2$ dimensional manifold.  
Those two sub-manifolds, however, are not generic; the upward flow ``seeks" values where the action is large and the thimble is bounded by the sub-manifold ${\rm det}D=0$ which makes their intersection an $N-1$ dimensional real manifold.}. 
In addition, the thimble is very curved near these singularities and frequently veers towards imaginary values of $\phi_i$. As the ``largest" values of $\phi_i$ (farthest from the critical point) are very far from the real values it attains asymptotically in the original integral over $\re^N$ it is doubtful that one thimble is enough to reproduce the desired integral. The analysis determining which combination of thimbles is equivalent to the original integration region $\re^N$ or, in other words, the determination of the integers $n_\sigma$ in \eq{eq:thimble_decomposition} is very involved in all but the simplest, one dimensional cases (for an example where this is possible, see~\cite{Kanazawa:2014qma}).

 
We will now describe the algorithm suggested in~\cite{Alexandru:2015xva}, referred to later on as the ``contraction algorithm", to sample one single thimble. It will be the basis of further generalization discussed later. We start by noticing that  every point $\phi$ sufficiently close to the critical point is indistinguishable from a point on the tangent plane to $\phi_c$. The tangent plane at the critical point can be characterized as the thimble corresponding to the quadratic part of the action. For this reason we will denote the neighborhood of $\phi_c$ which is  close enough to be approximated by the tangent plane the ``gaussian region". Taking a point in the gaussian region as the initial condition and evolving it according to~\eq{eq:flow_up} by a fixed time $T_{flow}$ establishes a mapping between the gaussian region and the thimble. By making $T_{flow}$ large enough we can guarantee that every point sampled on the thimble (or every point on the thimble if the thimble is finite) is mapped to a point close to the critical point. That establishes a mapping between the thimble and the gaussian region. Let us denote by $\tilde\phi$ the point in the gaussian region associated to a point $\phi$ on the thimble. The mapping between $\tilde\phi$ and $\phi$ has a jacobian $J$ that, as explained in~\cite{Alexandru:2015xva}, is the determinant of the matrix $J_{ij}(t=T_{flow})$ defined as the solution of
\bea\label{eq:flow_J}
 \frac{dJ_{ij}}{dt}  &=& \overline{  \frac{\partial^2 S[\phi]}{\partial\phi_i \partial\phi_k}J_{kj}  }, \\
 \frac{d\phi_i}{dt} &=& \overline{  \frac{\partial S[\phi]}{\partial\phi_i }  },
\eea  
with initial conditions $\phi(0)=\tilde\phi$ and  $J_{ij}(t=0)=e^{(i)}_j$, where $e^{(i)}_j$ form an orthonormal basis of the tangent space at $\phi_c$. The determinant $J$ contains information about not only  i) the inclination of the tangent space to the thimble at the point $\phi_i(T_{flow})$ with respect to the tangent space at $\phi_c$ (known in the literature as the ``residual phase") but also about ii)  the size of neighborhood of  $\phi_i(T_{flow})$ parametrized by a neighborhood of $\tilde\phi$ with unit volume. 
  
The contraction algorithm proposed in~\cite{Alexandru:2015xva} is a standard Metropolis algorithm in the variables $\tilde\phi$ using the effective action $S-\log |J|$. A little care has to be taken in the choice of update proposals, specially if $T_{flow}$ is very large as the appropriate values for the updates of $\tilde\phi$ are very different along different direction in field space. The algorithm becomes exact in the large $T_{flow}$ limit. In this limit, the sample points $\tilde\phi^{(a)}$ can be identified with points in the tangent space.

\section{$0+1$ dimensional Thirring model}
We will use the model studied in~\cite{Pawlowski:2014ada,Fujii:2015bua,Fujii:2015vha,Alexandru:2015xva} to illustrate our point. The model describes two flavors of fermions living on one single spatial site with a (continuum) action given by
\begin{equation}\label{eq:S-fermions}
L_{Th.}= \bar \chi \left (\gamma^0 {d \over dt} + m +\mu \gamma^0\right) \chi+ {g^2 \over 2}\left(\bar \chi \gamma^0 \chi \right)^2 \,,
\end{equation}
where $\chi$ is a two component, time dependent spinor and $\gamma^0$ is a Pauli matrix. The interaction term is simply the 0+1 dimensional analog of the current-current interaction $(\bar\chi\gamma^\mu\chi)(\bar\chi\gamma_\mu\chi)$ of the  $1+1$ dimensional Thirring model. The quartic interaction can be eliminated by introducing an auxiliary field $\phi$ and, upon integrating over $\bar\chi, \chi$ the partition function is
\beq
Z=\int {\cal D} \phi\,\det\left(\gamma^0 {d \over dt} + i \gamma^0 \phi +  m +\mu \gamma^0 \right) e^{- {1\over 2 g^2} \int dt  \phi^2 }
=
\int {\cal D} \phi\,
e^{- {1\over 2 g^2} \int dt  \phi^2 -{\rm tr} \log(D[\phi])}
\, ,
\eeq 
with $D[\phi] =\gamma^0 {d \over dt} + i \gamma^0 \phi +  m +\mu \gamma^0  $. This model has  a sign problem for non-zero values of $\mu$ and it  has been used as a toy model for testing ideas such as complex Langevin dynamics~\cite{Aarts:2008rr,Pawlowski:2014ada,Pawlowski:2013pje} and hybrid Monte Carlo on Lefschetz thimbles~\cite{Fujii:2015vha,Fujii:2015bua}.

\bigskip\label{fig:quenchedphase}
\begin{figure}[!tbp]
  \centerline{\includegraphics[height=5.2cm]{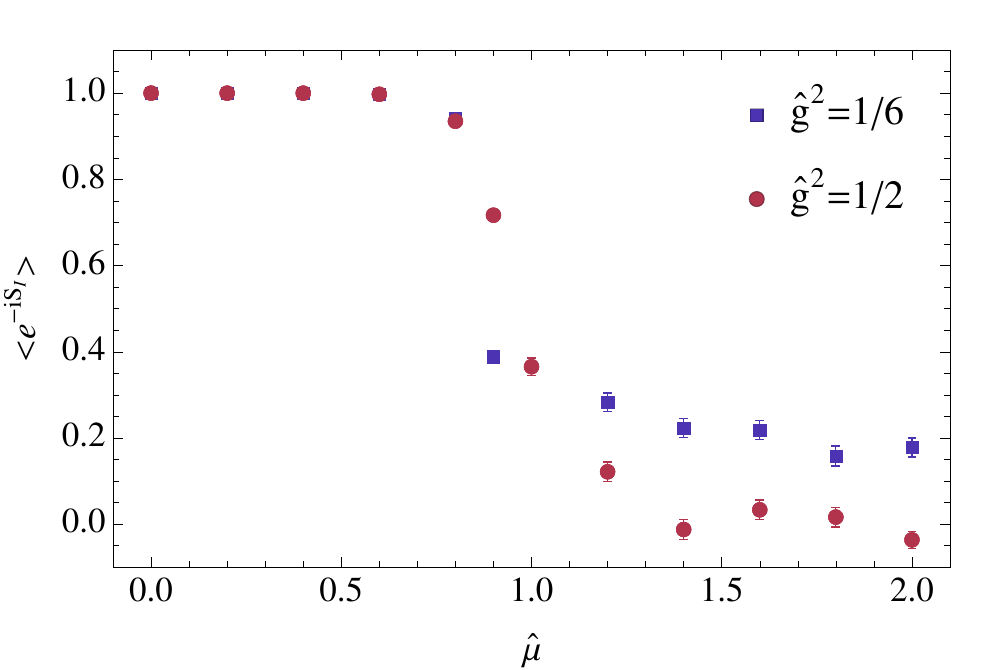}}
\noindent
\caption{Average phase $\langle e^{-i S_I}\rangle_R$ in the reweighting method for $N=8, \hat m=1$ and $\hat g^2=1/6$ (blue) and $\hat g^2=1/2$ (red).}
\label{fig:quenchedphaseplot}
\end{figure}  

Discretizing the (euclidean) time coordinate and using a staggered fermion formulation we obtain the lattice model defined by
\begin{equation}
Z=\left[\prod_{t=1}^{N} \int_{0}^{2\pi} {d \hat\phi_t\over 2\pi}\right] \det D e^{-{1\over 2\hat g^2} \sum_{t=1}^N (1-\cos \hat\phi_t)}\equiv\left[\prod_{t=1}^{N} \int_{0}^{2\pi} {d \hat\phi_t\over 2\pi}\right] e^{-S[\hat\phi]}, \label{eq:Z-lattice}
\end{equation}
where the effective action and the explicit form of the discretized Dirac matrix are
\begin{eqnarray}\label{eq:S-bosons}
S[\hat\phi]&=&{1\over 2\hat g^2} \sum_{t=1}^N (1-\cos \hat\phi_t)-\log\det D[\hat\phi] \,,\label{eq:S} \\
D_{t,t^\prime}[\hat\phi]&=&{1\over 2}\left(e^{\hat\mu+i \phi_t}\delta_{t+1,t^\prime}-
e^{-\hat\mu-i \hat\phi_{t^\prime}}\delta_{t-1,t^\prime}+
e^{-\hat\mu-i \hat\phi_{t^\prime}}\delta_{t,1}\delta_{t^\prime, N} -
e^{\hat\mu+i \hat\phi_{t}}\delta_{t,N}\delta_{t^\prime, 1}\right)+\hat m\,\delta_{t,t^\prime}\,. \label{eq:D}
\end{eqnarray}
Here $N=\beta/a$ is an even number that denotes the number of lattice sites related to 
the inverse temperature of the system $\beta$, and all the dimensionful quantities, 
$m$, $g^2$, $\mu$ are rendered dimensionless  by multiplying with appropriate 
powers of the lattice spacing: $\hat m=ma$, $\hat\mu = \mu a$, $\hat g^2=g^2 a$, $\hat\phi=a\phi$. 
The auxiliary field $e^{i\hat\phi}$ in this discretization plays the role of a $U(1)$ link variable. 

We can gauge the severity of the sign problem by the average phase $\langle e^{-i S_I}\rangle$ in a standard Metropolis calculation with real fields $\hat\phi$. \fig{fig:quenchedphaseplot} shows the average phase for two values of the coupling strength $\hat g^2$ and $\hat m=1$. For parameters where the average phase is much smaller than one, a Monte Carlo evaluation requires a large number of configurations and is impractical in any realistic model.

The partition function, the chiral condensate, and the charge density can be calculated 
analytically in this lattice model~\cite{Pawlowski:2014ada} and are useful in assessing the accuracy of the numerical calculation:
\begin{eqnarray}\label{eq:exact}
Z&=&{e^{-N\alpha}\over 2^{N-1}}\left[I_1^N(\alpha)\cosh(N\hat\mu)+I_0^N(\alpha) 
\cosh(N \sinh^{-1}(\hat m))\right] \,,\\
\langle n \rangle&=&{1\over \beta}{\partial \over \partial \mu} \log Z=
{1\over N}{\partial \over \partial \hat\mu} \log Z= {I_1^N(\alpha)\sinh(N\hat\mu) \over I_1^N(\alpha)\cosh(N\hat\mu)+I_0^N(\alpha) \cosh(N \sinh^{-1}(\hat m)) } \,,\\
\langle \bar\chi\chi \rangle&=&{1\over \beta}{\partial \over \partial m} \log Z=
{1\over N}{\partial \over \partial\hat m} \log Z= {(1+\hat m^2)^{-1/2}I_0^N(\alpha)\sinh(N\sinh^{-1}(\hat m)) \over I_1^N(\alpha)\cosh(N\hat \mu)+I_0^N(\alpha) \cosh(N \sinh^{-1}(\hat m)) }\,,
\end{eqnarray} 
where $\alpha\equiv1/(2 \hat g^2)$ and $I_n(\alpha)$ denotes the modified Bessel function of the first kind. 

\begin{figure}[t]
\includegraphics[height=5.2cm]{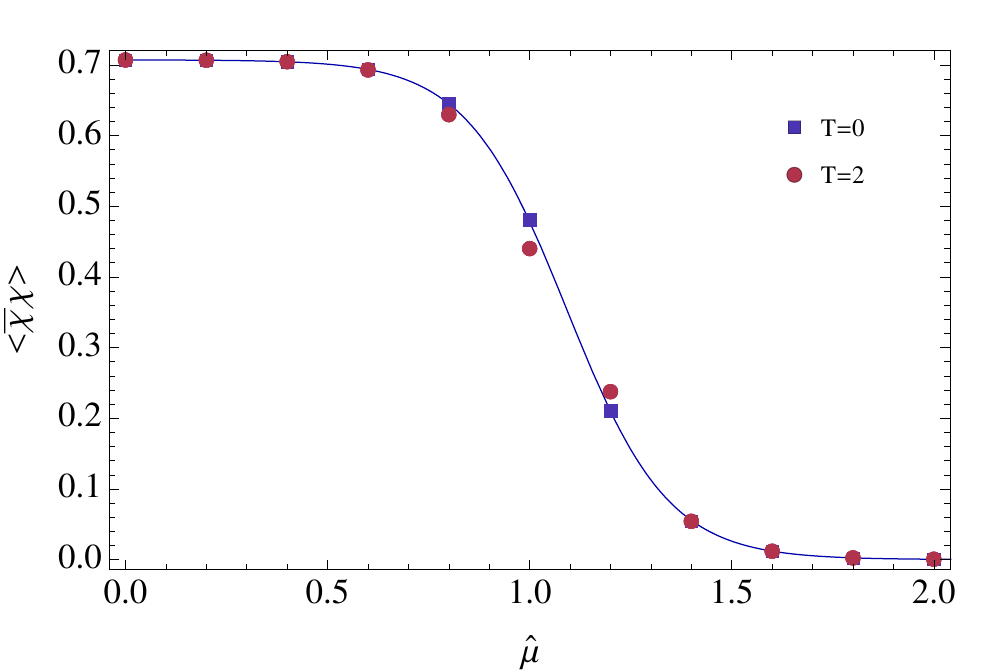}
\hskip 1.2cm
\includegraphics[height=5.2cm]{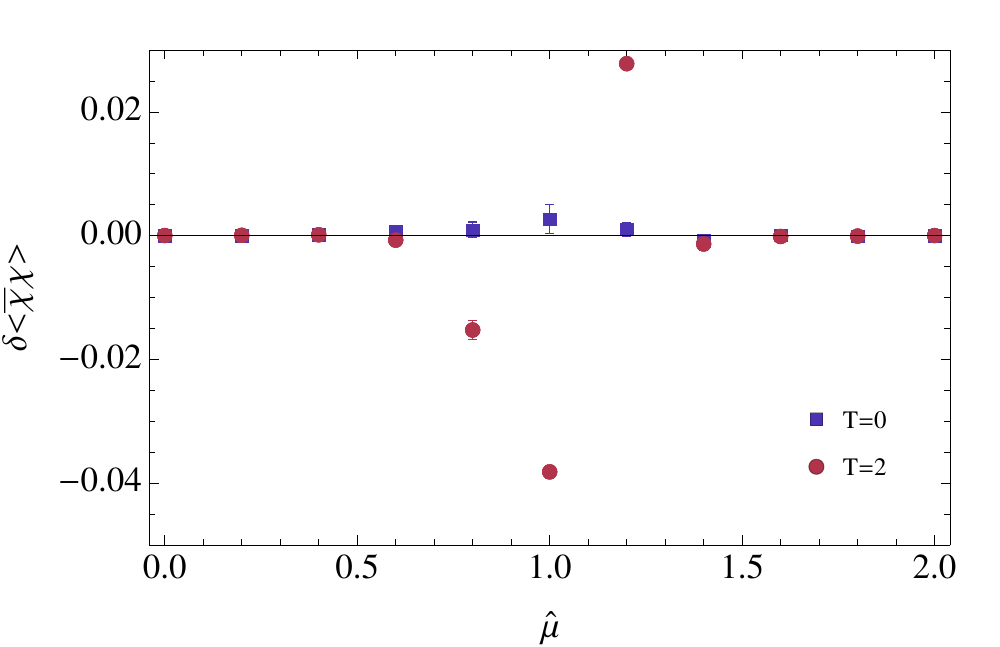}
\caption{Condensate value as a function of chemical potential $\hat\mu$ (left) and difference from the exact condensate value (right). The parameters are $N=8, \hat m=1, \hat g^2=1/6$. The blue points are the result of the integration over the tangent space while the red points were obtained in~\cite{Alexandru:2015xva} by using the contraction algorithm with flow time $T_{flow}=2$ and, essentially, correspond to the integration over the main thimble.}
\label{fig:N8beta3}
\end{figure}

The critical points of action in \eq{eq:S-bosons} are determined by the equation
\beq\label{eq:gap_equation}
0=\frac{\partial S}{\partial\hat\phi_t} = \frac{1}{2\hat g^2}\sin(\hat\phi_t)
 + \frac{\sin(\sum_{t'}\hat\phi_{t'}-i N\hat\mu)}{\cos(\sum_{t'}\hat\phi_{t'}-i N\hat \mu)+M}
\eeq 
with $M=(\hat m+\sqrt{1+\hat m^2})^N + (\hat m-\sqrt{1+\hat m^2})^N$.
The solution to \eq{eq:gap_equation} with  the smallest (real part of the) action has the form $\hat\phi_t = (i \bar\phi, \cdots, i \bar\phi)$ with $\bar\phi$ real. The tangent space to this point is spanned by purely real vectors, a feature common in this class of models. As we will see below this critical point and its associated thimble and tangent space will play a central role in our discussion and for this reason will be called ``main thimble" and ``main tangent space". 
In~\cite{Alexandru:2015xva} the contraction algorithm discussed above was applied to the computation of $n$ and $\langle \bar\chi \chi\rangle$ over the main thimble. The results showed small but statistically significant  discrepancies from the exact result (\eq{eq:exact}) for some values of the parameters. For instance, in~\fig{fig:N8beta3} we show the condensate as a function of $\hat\mu$ for the parameters $\hat g^2=1/6, N=8, \hat m=1$ where the discrepancy is evident. Similar results were also obtained in~\cite{Fujii:2015vha}. 
There is evidence that the discrepancy between Monte Carlo and exact results are due to the neglected contributions of other thimbles. In fact, a semiclassical estimate of the contribution of the other thimbles, which is suppressed by their higher action, suggests that they have the right order of magnitude to explain the discrepancies with the exact result. This is, however, only a plausibility argument as the semiclassical expansion is not reliable for the parameters considered\footnote{The semiclassical expansion is an expansion in power of $\hat g^2/N_F$, where $N_F=1$ is the number of fermion flavors.} and the possibility that the discrepancy arises from a problem in the algorithm computing the contribution of the main thimble should not be excluded.

\section{Beyond Thimbles}

We have up to now described the contraction algorithm as a way to compute the integral over the main thimble. It becomes exact in the limit of large flow time ($T_{flow}\rightarrow\infty$) but, as we will  argue now,   exact results can also be obtained by using the contraction algorithm at {\it finite} values of $T_{flow}$, which is profitable to do in many circumstances.

Let us first consider the case where  field variables are unbounded  and the action $S$ is a polynomial which diverges only at infinity.
Since the integrand in the partition function is holomorphic, the manifold of integration can be deformed without changing the partition function ~\cite{shabat1992introduction}. But there are restrictions on the allowed deformations. For the path integral to be well defined it is necessary for the real part of the action $S_R$ to diverge at large values of the field. Otherwise the integrand does not approach zero asymptotically and the integral does not converge. The space of ``large fields'' where $e^{-S_R(\phi)}<\epsilon$, which ensures the converge of the integral, is in general a disjoint union of submanifolds. Therefore there are multiple different integration manifolds (over which the path integral is convergent) with different asymptotic regions taking values in this disjoint set. These manifolds are separated into discrete homology classes such that two such manifolds over which the integral has the same value, belong to the same class \cite{fedoryuk,pham}. In other words, different homology classes differ in the asymptotic behavior of their manifolds. We will argue now that the (upward) flow does not change the homology class of a manifold. Let us take the points of a manifold $M(0)$ as the initial condition of the flow equations. After evolving each point of $M(0)$  by a fixed time $T_{flow}$ we obtain another manifold, $M(T_{flow})$.  Since the homology classes form a discrete set,  in order for $M(T_{flow})$ to be in a different homology class it is necessary that at some intermediate flow time $t<T_{flow}$ the integral over $M(t)$ be ill defined. 
But the integral over $M(t)$ is well defined at all $t$. Indeed,  in order for  the integral over $M(t)$ not to be defined there has to be a path to infinity $L_t$ on $M(t)$ over which $S_R$ is bounded from above. Consider then the pre-image $L_0$ of $L_t$ under the flow by $t$. Since the flow only increases the value of $S_R$, the values which $S_R$ takes along $L_0$ would also be bounded and the integral over $M(0)$ would not be well defined, which is contrary to our assumption. Consequently, $M(T_{flow})$ is in the same homology class as $M(0)$. 

  In the fermionic model we are considering, the action is not a polynomial and two small modifications are needed to the argument we just made. The first is that the real parts of the field variables $\phi$ are bounded. The second one is that $S_R$ diverges at finite values of the field and we consider finite manifolds that end on these singularities. It is the behavior of $S_R$ as these points are approached that determine the homology class of the manifolds. The points where $S_R$ diverges play the role that points at infinity play in the bosonic  (polynomial $S$) case.

 In the present model (and similar ones) the main tangent space is purely real, and thus parallel to the original domain of integration $\mathbb{R}^N$. Consequently, the main tangent space has the same asymptotic behavior and belongs to the same homology class as  $\mathbb{R}^N$. Now, consider the manifold $M(T_{flow})$ obtained by flowing the main tangent space along the upward flow by a flow time $T_{flow}$. As argued above, $M(T_{flow})$ is also in the same homology class as the main tangent space $M(0)$. As $T_{flow}$ increases, the evolution of some special points of $M(0)$ approaches  critical points of the action. Points in $M(0)$ near those special points will flow towards points close to the thimbles associated with each one of these critical points. 
 As the flow is tangent to the thimbles it cannot cross a thimble but, instead, approaches it asymptotically.
 Other points in $M(0)$  flow towards infinity or, in non-polynomial actions, to singularities of $S$. The result is that $M(T_{flow})$ will approach all the thimbles contributing to the original integral, and only those. If that was not the case, the integral over $M(T_{flow})$ for large $T_{flow}$ would have a different value from the one over $M(0)$, in contradiction to the argument that they belong to the same homology class. This argument is one way of understanding the  rule determining the coefficients $n_\sigma$ in 
 \eq{eq:thimble_decomposition} stating that $n_\sigma$ are determined by the intersection of the downward set of the thimble $\sigma$ with the original domain of integration.  For example, if 2 points on $M(0)$ flow to the critical point of thimble $\sigma$, then $n_\sigma$ is 2.  If the flow transports a tangent basis attached to one of these special points to an oppositely oriented basis, then $n_\sigma$ is negative (or the integral over thimble $\sigma$ is defined to have an opposite orientation).
 
\begin{figure}[t]
\includegraphics[height=6.2cm]{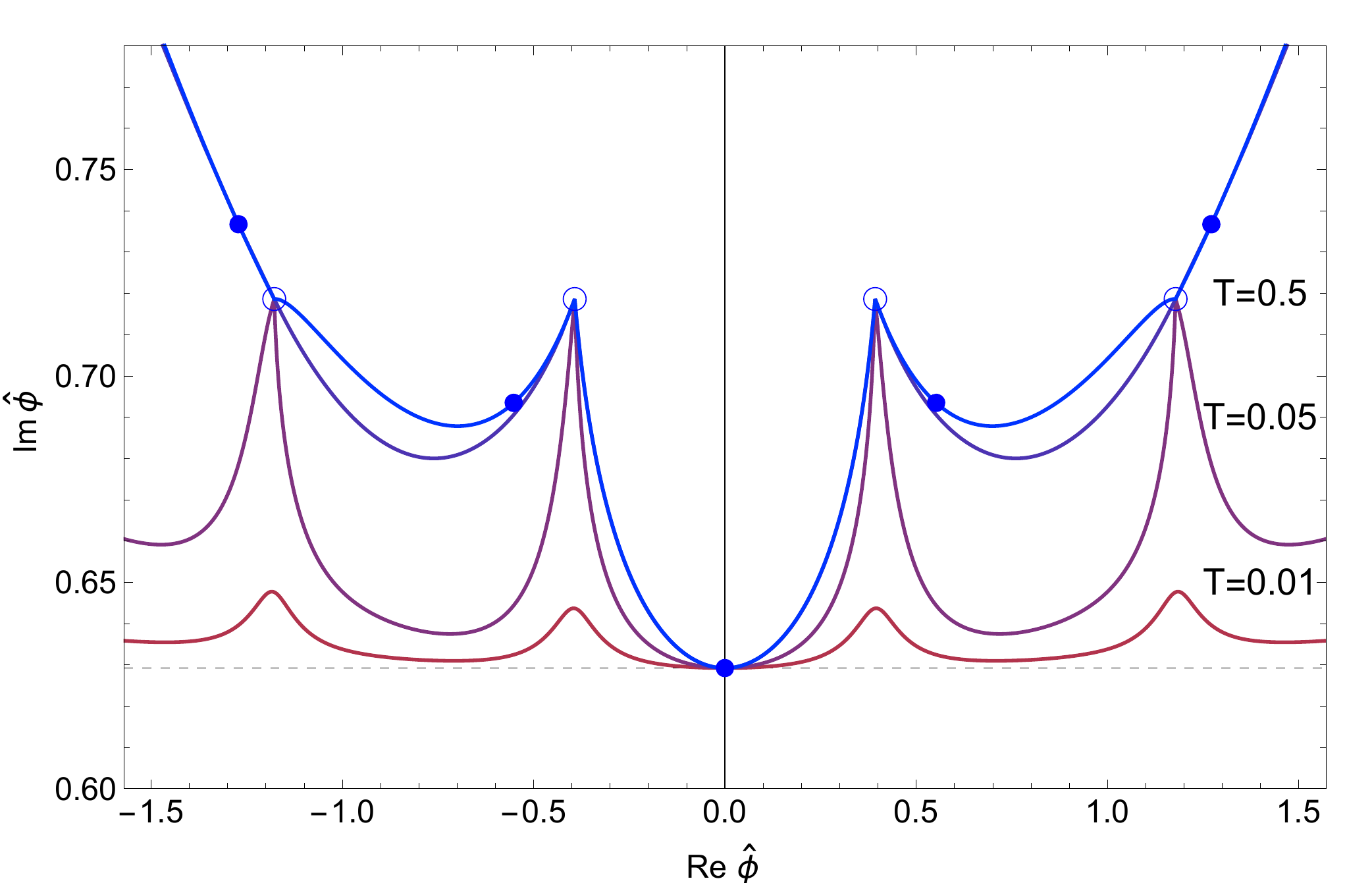}
\caption{Complex $\hat\phi=\frac{1}{N}\sum_t \hat\phi_t$ plane. The full (blue) circles are the critical points of $S$, the empty (blue) circles the log singularities of $S$ (points where the fermion determinant vanishes).  The full (blue) line are five of the thimbles. The red dashed line is the main tangent space.  The red, purple and violet lines are the result of flowing the tangent space by $T_{flow}=0.01, 0.05$ and $0.5$, respectively.}
\label{fig:bigpicture}
\end{figure}

\begin{figure}[t]
\includegraphics[height=4.2cm]{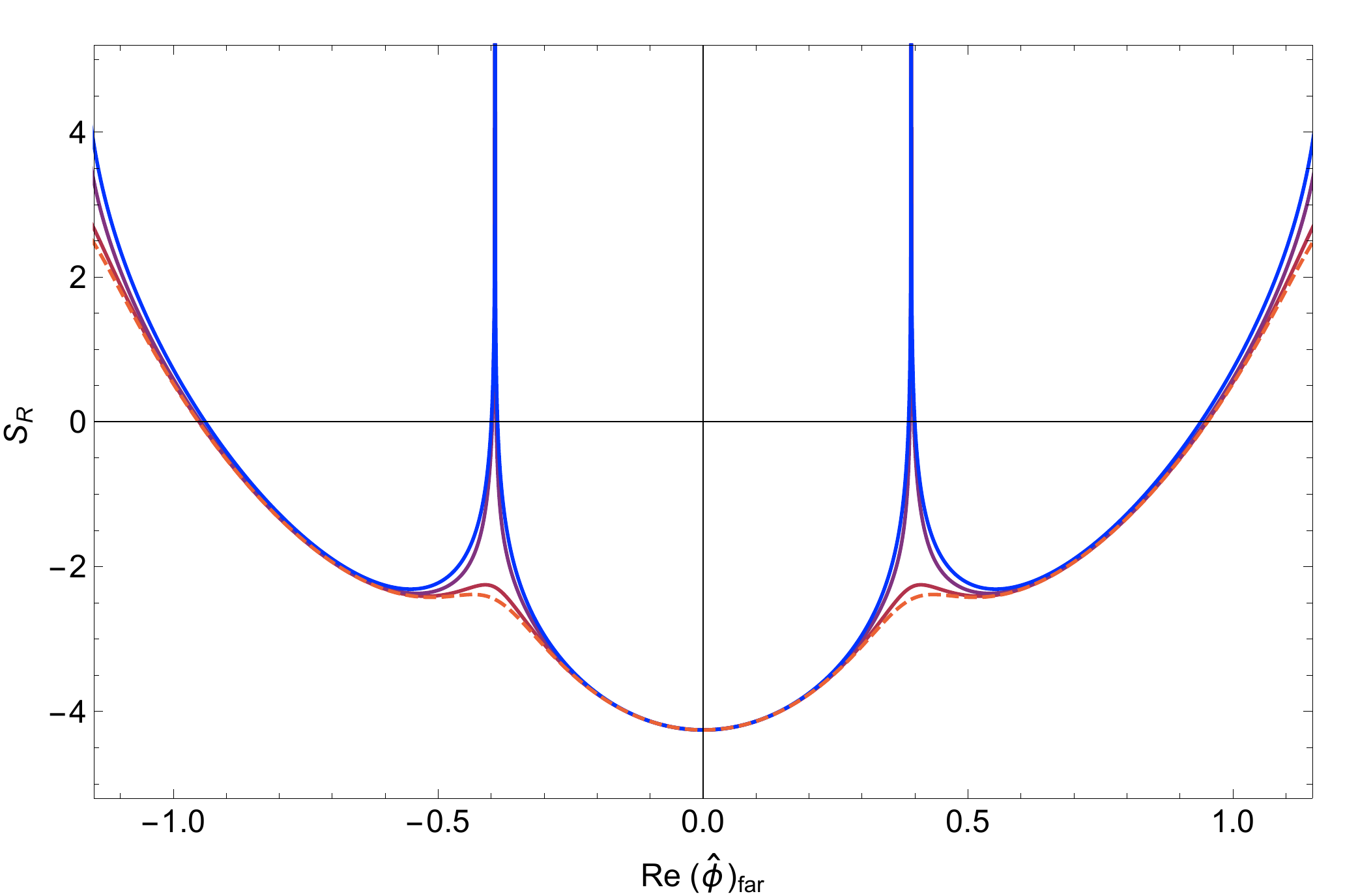}
\hskip 1.2cm
\includegraphics[height=4.2cm]{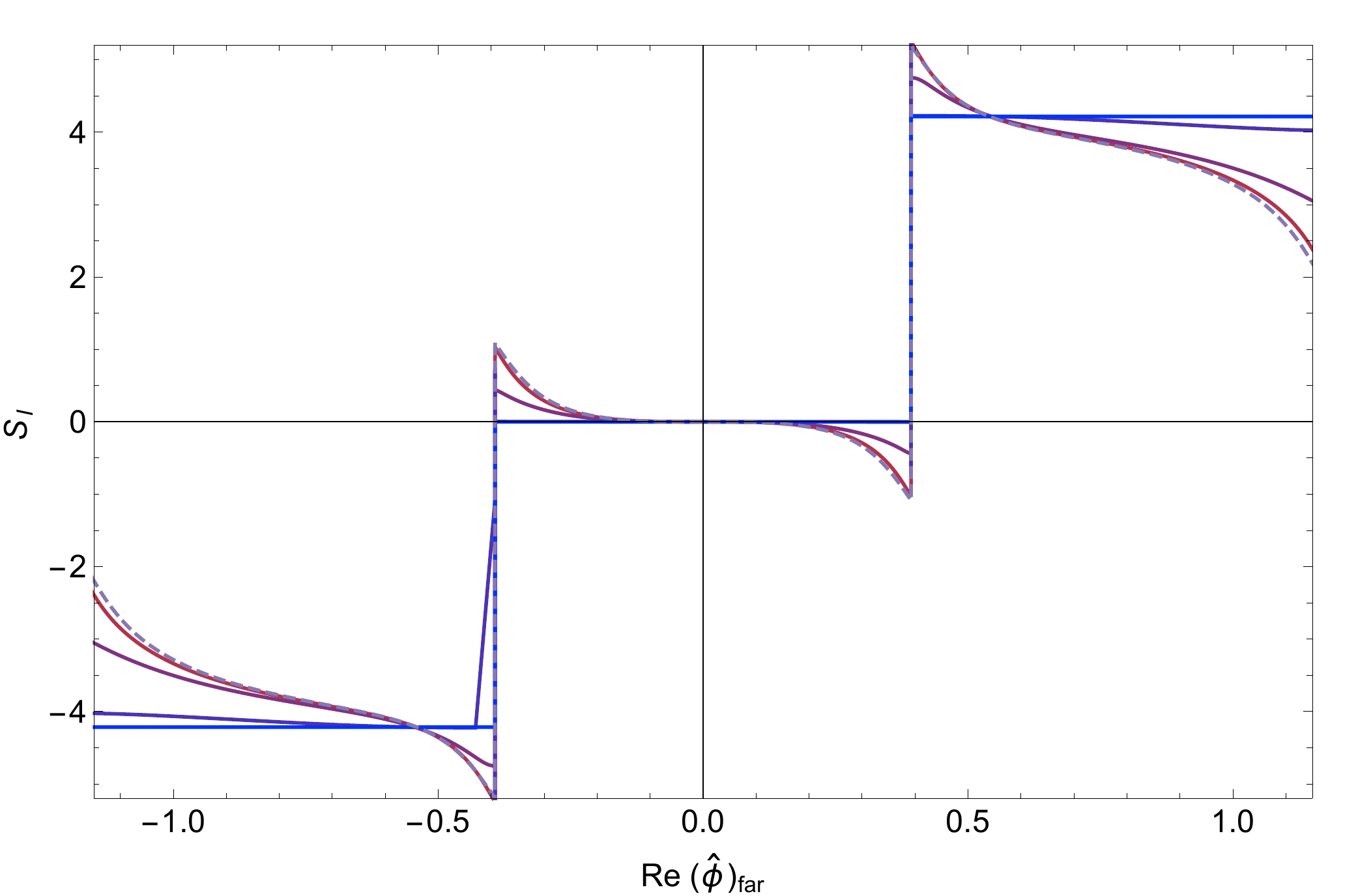}
\caption{
Real (left panel) and imaginary (right panel) parts of the action along the curves shown in \fig{fig:bigpicture}.  As the flow increases the imaginary part of the action becomes more (piecewise) constant while the barriers on the real part become more prominent.
}
 \label{fig:actionmelting}
\end{figure}

   The observations above suggest a practical way of performing the original integral even in the case where many thimbles contribute and even if no direct knowledge of the location of the critical points is available. 
   The contraction algorithm with $T_{flow}=0$ computes the integral over the main tangent space
   that, as mentioned above, is  equivalent to the integral over $\mathbb{R}^N$. However, a severe sign problem may still exist on the  main tangent space and a way to ameliorate that is to use a manifold obtained by flowing by a finite amount $T_{flow}$. In $M(T_{flow})$ the regions with small value of  $S_R$ -- which dominate the integral -- are smaller than in $M(0)$. On the other hand the imaginary part $S_I$ is the same in corresponding points of $M(T_{flow})$ and $M(0)$ (since the flow preserves $S_I$). Thus, the regions of $M(T_{flow})$ with large statistical weight have a smaller phase fluctuation and the sign problem is reduced in $M(T_{flow})$. In the limit of large flow time $T_{flow}$, a set of small regions in the main tangent space are mapped by the flow into manifolds approaching each of the relevant thimbles (the ones that contribute to the original integral). The regions that are not in this set, flow to regions where $S_R$ approaches infinity. Also,  $S_I$ in each of these patches in this set is approximately constant since at large $T_{flow}$ the regions are small and $S_I$ cannot vary much over them. The whole picture is exemplified in Figs. \ref{fig:bigpicture} and \ref{fig:actionmelting}. The mapping between the main tangent plane and $M(T_{flow})$ has some confusing features so it is worthwhile picturing an example. \fig{fig:bigpicture} was generated  with  the parameters $N=8, \hat m=1,\hat \mu=1.6, \hat g^2=1$) and we projected the $N=8$ dimensions onto the complex plane of the average field $\hat\phi=\frac{1}{N}\sum_t\hat\phi_t$ for better visualization. As explained above, the thimbles (blue lines) are joined by the points where the fermion determinant vanishes and the action has a singularity. By increasing the flow time ($T_{flow}=0.01, 0.05$ and $0.5$) the tangent space is transported closer to the location of the thimbles. Thimbles that do not contribute to the partition function (not shown in the figure) are not approached by flowing from the tangent space. 
%

  All algorithms previously proposed  for the integration over one thimble require the corresponding critical point to be known analytically. In a more realistic field theory, the integration over all relevant thimbles would require that {\it all} solutions of the (euclidean) equation of motion, {\it even with complexified fields}, be known. Furthermore, one would have to determine which one of those contribute to the original integral over real fields, a daunting task that has only been accomplished in $0+0$ and $0+1$ dimensional models. 
  A main point of this paper is that  the observations of this section imply that the contraction algorithm, used at finite values of $T_{flow}$, is an alternative way of computing the original integral {\it without assuming the dominance of a single thimble}.
  In other words, as the points of the tangent space ($\tilde\phi$) parametrize (through the flow) a manifold in the same homology class as $\mathbb{R}^N$, a 
 Markov chain in the tangent space  generated with the effective action $S-\log |J|$ will sample {\it all} relevant thimbles, bypassing the need to analytically find all critical points and their corresponding coefficients $n_\sigma$.



Let us now illustrate the points above with explicit calculations.
The integral over the main tangent space is extremely cheap computationally as the flow equations \eq{eq:flow_eq} do not need to be solved (the equations for  the matrix $J_{ij}$ are by far the most expensive part of the algorithm). Of course, the integrand is not real over the main tangent space and a potential sign problem arises. Still, for some models/parameter sets the phase $e^{-i S_I}$ can be reweighted {\it even in cases when a similar reweighting on the original domain of integration is not feasible}. An example of this is shown in \fig{fig:N8beta3} where the result of the calculation in~\cite{Alexandru:2015xva}  (corresponding to the contribution of the main thimble only) is plotted together with a similar calculation performed with flow $T_{flow}=0$ (corresponding to an integration over the tangent space $M(0)$). 
 
\begin{figure}[t]
\includegraphics[height=5.2cm]{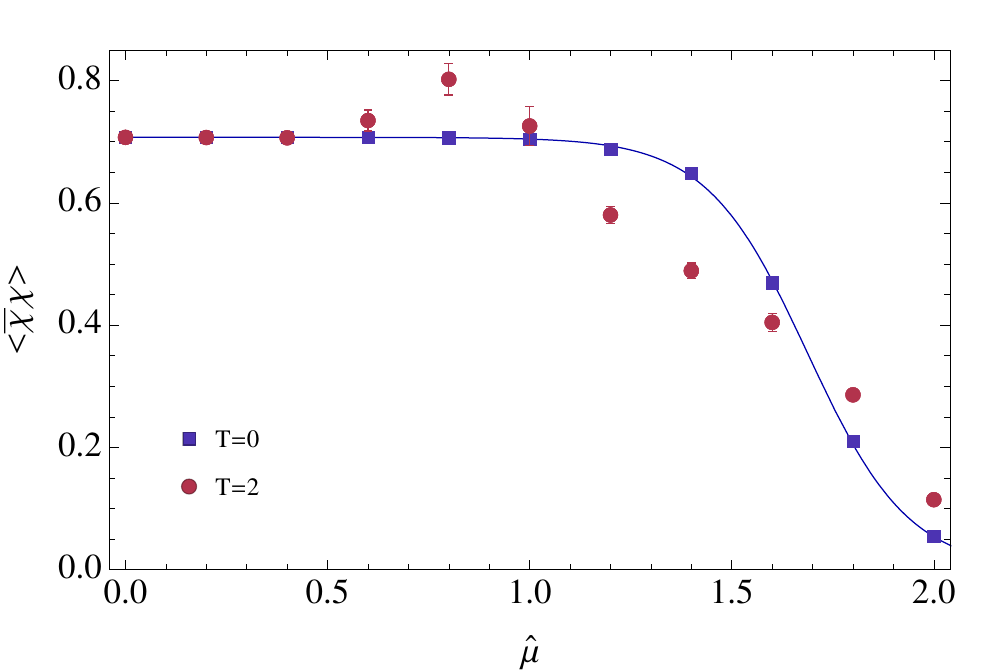}
\hskip 1.2cm
\includegraphics[height=5.2cm]{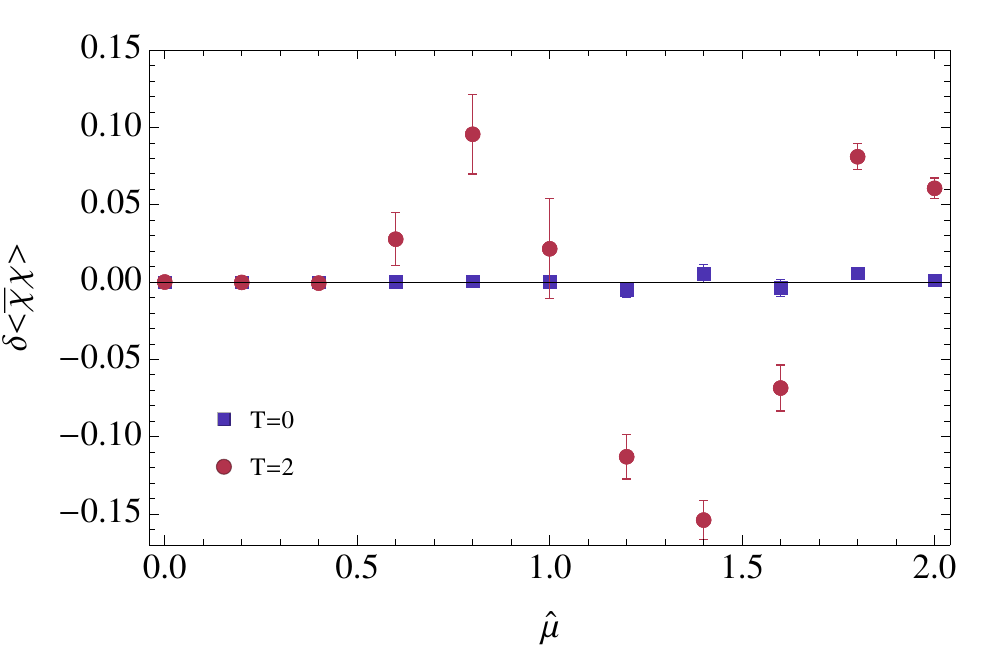}
\caption{Condensate value as a function of chemical potential $\hat\mu$ (left) and difference from the exact condensate value (right). The parameters are $N=8, \hat m=1, \hat g^2=1/2$. The blue points are the result of the integration over the tangent space while the red points were obtained in~\cite{Alexandru:2015xva} by using the contraction algorithm with flow time $T_{flow}=2$ and, essentially, correspond to the integration over the main thimble.}
\label{fig:N8beta1}
\end{figure}

The possibility of reweighting the $e^{-i S_I}$ phase on the main tangent space is not an artifact of small coupling. In the strong coupling $N=8, \hat m=1, \hat g^2=1/2$ case, where the quenched phase $\langle e^{-i S_I}\rangle_R$ essentially vanishes beyond $\hat\mu\approx 1.4$, the $e^{-i S_I}$ phase is not smaller than $\approx 0.3$ in the main tangent plane and is easily reweighted. As  shown in~\fig{fig:N8beta1} the results for the condensate agree with the exact result. On the other hand,  the $T_{flow}=2$ flow result does not agree with the exact result. In view of our discussion above we can see why it fails.
The contraction algorithm computes the integral over the manifold $M(T_{flow})$ obtained by the upward flow of the main tangent space. As $S_R$ is increased by the flow, the region of the tangent space that is mapped into a manifold close to the main thimble becomes  separated from the regions mapped into other thimbles by broader and taller action barriers. Our algorithm, based on a  Metropolis chain in the tangent space using the effective action $S_R-\log|J|$, ``gets stuck" in a small region mapped into the main thimble. Consequently, only the main thimble is sampled. In fact, the $T_{flow}=2$ results
agrees with~\cite{Fujii:2015vha} obtained with a different algorithm sampling (by construction) only the main thimble.
These results demonstrate that i) the disagreement between the results in~\cite{Alexandru:2015xva,Fujii:2015vha} and the exact result come indeed from the neglected contribution from other thimbles and ii) the contribution from other thimbles are captured by the $T_{flow}=0$ no flow contraction algorithm which computes the integral over the main tangent space.
 
Given the success of the zero flow contraction algorithm in dealing with system with small $N$, a natural question is how well this approach does with larger $N$. $N$ can be increased along two different routes of physical interest: the continuum limit and the zero temperature limit. 

\begin{figure}[t]
\includegraphics[height=5.2cm]{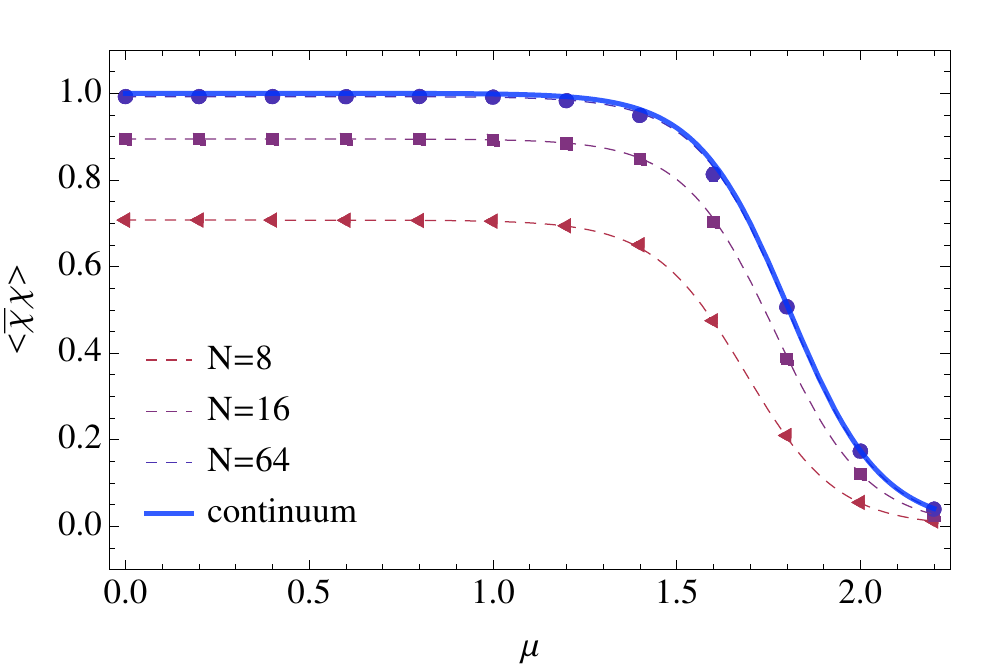}
\hskip 1.2cm
\includegraphics[height=5.2cm]{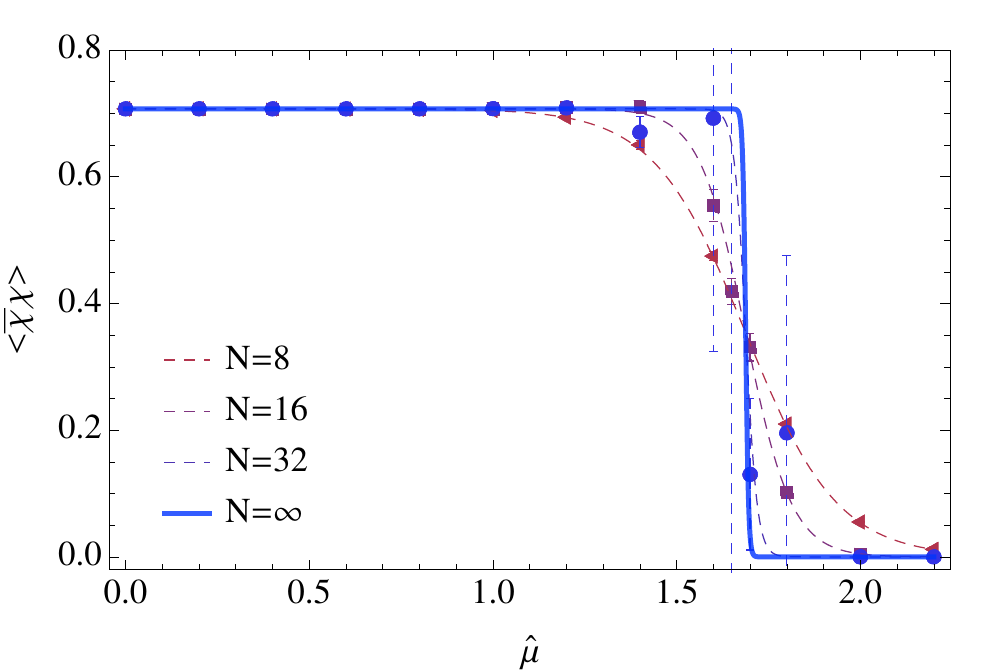}
\caption{Condensate value as a function of chemical potential $\mu$ in the continuum limit (left) and 
in the zero temperature limit (right). The exact result is shown in both cases as a continuous (blue) line. 
The parameters for the $N=8$ case are $\hat m=1, \hat g^2=1/2$, corresponding to the strongly coupled case.
For other values of $N$ the remaining parameters are scaled as explained in the main text.}
 \label{fig:extrapolations}
\end{figure}

 Consider first the continuum limit. It corresponds to the limit $N\rightarrow\infty$ while keeping and $\hat g^2N, \hat mN, \hat \mu N$ fixed. The results from the zero flow calculations, that is, the result of the integration over the main tangent space, are shown in the left panel of \fig{fig:extrapolations} starting from the strongly coupled (more challenging) case $N=8, \hat g^2=1/2, \hat m=1$. It is evident that, as expected,  the continuum limit is approached smoothly and no problem arises due to the fluctuating $e^{-i S_I}$ phase. On the other hand it is expected that the $e^{-i S_I}$ phase will fluctuate more and more as the zero temperature is approached since the euclidean action is proportional to the spacetime volume and, consequently, proportional to the inverse temperature. This is indeed what we find. In the right panel of~\fig{fig:extrapolations} we show the results of the zero flow computation starting from the $N=8,\hat g^2=1/2,\hat m=1$ case and taking the limit $N\rightarrow\infty$ while keeping $\hat g$, $\hat m$ and $\hat\mu$ constant. For large enough $N$ (low enough temperature) the $e^{-i S_I}$ phase fluctuates too wildly to be reweighted without resorting to enormous numbers of configurations (which could be done in our toy model but would be impractical in problems of greater physical interest). We stress, however, that by integrating over the main tangent space one can perform calculations at much lower temperatures than one could have by integrating over real fields.
 
The examples above illustrate the best case scenario: a simple change of variable ($\hat\phi\rightarrow \hat\phi + \hat\phi_c$) is enough to make the sign problem tractable by reweighting. From these examples  we know this situation is more likely to happen at weak coupling. On the other hand, the action is an extensive quantity and we do expect the phase $e^{-i S_I}$ (calculated for real variables or on the main tangent space) to vanish in the thermodynamic limit. It is unclear how the phase behaves on the main tangent space in asymptotically free theories in the continuum (small coupling) and thermodynamic/zero temperature limits. Whether the integration over the main tangent space alleviates the sign problem enough to be useful to physically interesting problems is something that can be determined only by future  calculations.
  

In order to tame the sign problems at low temperatures we apply the algorithm with a finite flow time $T_{flow}$. There is a compromise in the choice of $T_{flow}$. Too small a value does not alleviate the sign problem enough. For too large values of $T_{flow}$ the relevant regions in the tangent space (which parametrizes the important parts of the thimbles) are separated by regions of large $\mathbf{Re}(S-\log|J|)$ which ``trap" the Markov chain and leads to large decorrelation lengths.  
At infinite $T_{flow}$ these barriers correspond to the junction of thimbles at their borders  where $S_R$ diverges.
That is why our previous calculation in~\cite{Alexandru:2015xva}, performed with a fairly large value of $T_{flow}$, was constrained to the region of the tangent space mapped onto the main thimble,  failed to include the contributions corresponding to the remaining thimbles and produced some wrong results corresponding to the contribution of only one thimble (typically, at larger couplings)\footnote{The contribution of one thimble was properly computed as the agreement with the results in ~\cite{Fujii:2015vha,Fujii:2015bua} indicate. }. 

\begin{table}[t!]
  \centering
\begin{tabular}{|c|c|c|}
\hline
$N$ &   $\langle \bar\chi \chi\rangle$   &    $\langle \bar\chi \chi\rangle_{exact}$ \\
\hline
$16$    &  0.330(13)    &    $0.353$    \\
\hline
$32$ & $0.345(46)$ & $0.353$ \\
\hline
$64$     &   0.375(48)     &    $0.353$      \\
\hline
\end{tabular}
\caption{Value of the condensate in the low temperature ($N\rightarrow\infty$) limit and $\mu=1.688$ obtained with  flow time $T_{flow}=0.5$. These results and its error bars should be compared to the ones in the right panel of \fig{fig:extrapolations} obtained with $T_{flow}=0$. }
  \label{tab:lowTcriticalmu}
\end{table}

\begin{figure}[t]
\includegraphics[height=3.2cm]{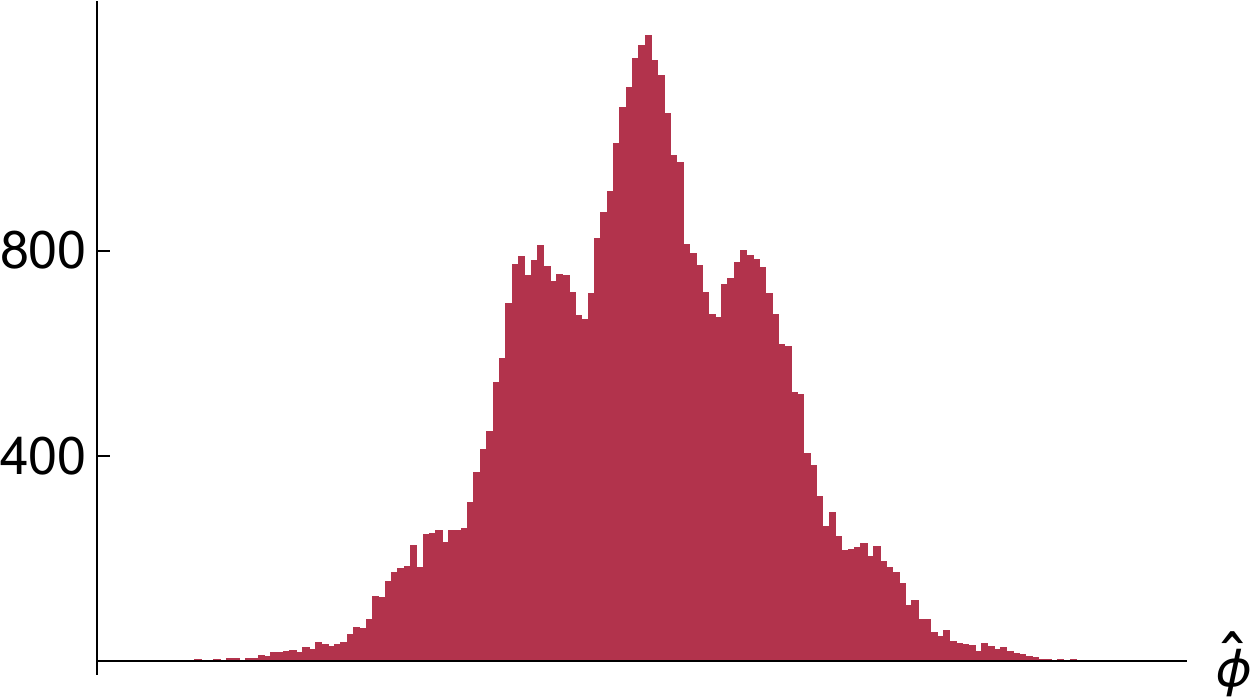}
\hskip 1.2cm
\includegraphics[height=3.2cm]{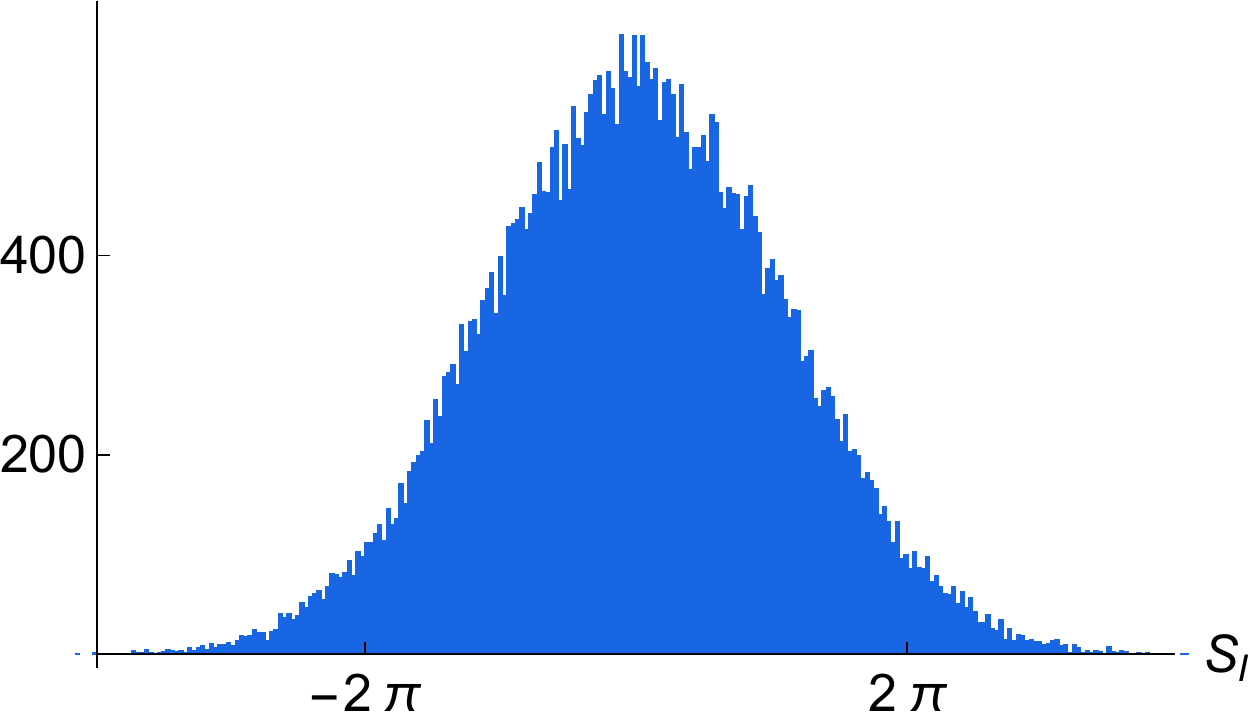}
\vskip 1cm
\includegraphics[height=3.2cm]{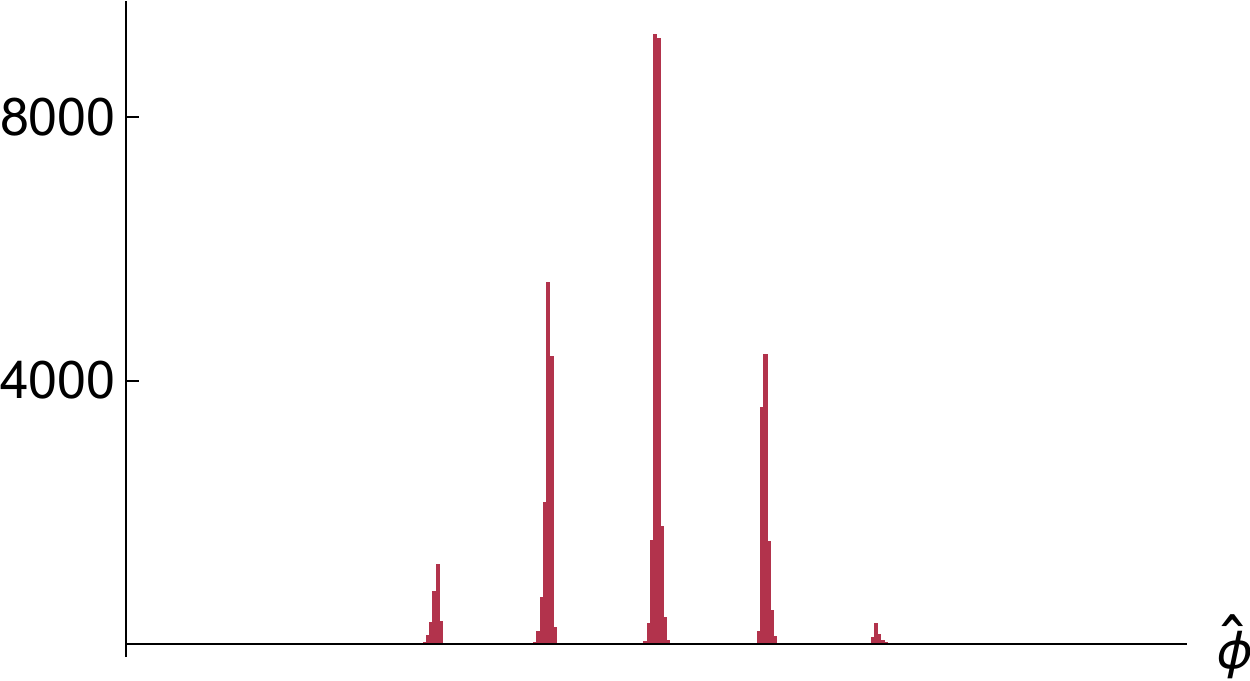}
\hskip 1.2cm
\includegraphics[height=3.2cm]{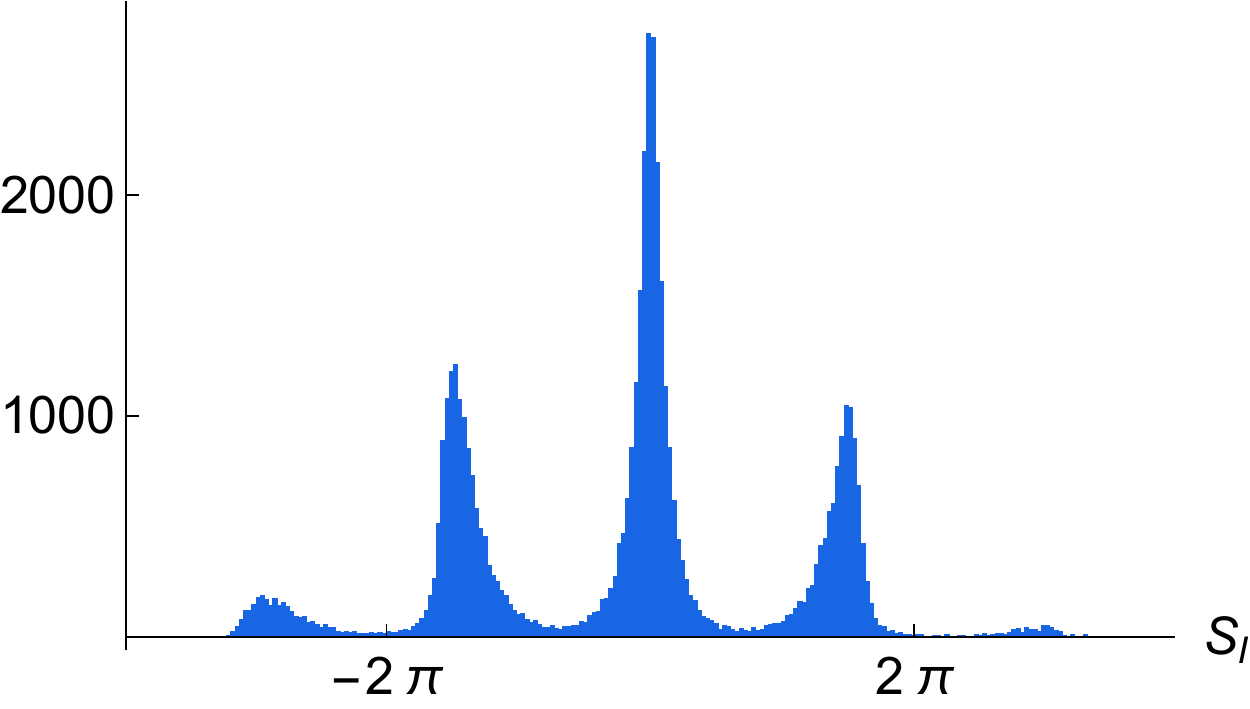}

\caption{Monte Carlo sampling  in a $T_{flow}=0$ (top line) and  $T_{flow}=0.5$ (bottom line), N=32 calculation. The left plot shows the average value $\hat\phi=\sum_t \hat\phi_t/N$ of the fields. The right panel shows the imaginary part of the action of the flowed field $S_I(\phi)$. 
In the $T_{flow}=0$ calculation on the top line the phase $e^{-i S_I}$ fluctuates too wildly and the result is the large error bars in \tab{tab:lowTcriticalmu}. On the $T_{flow}=0.5$ calculation on the bottom line the phase fluctuates much less. It is also evident that regions on the the tangent space corresponding to several thimbles are being sampled.}
 \label{fig:histograms}
\end{figure}

After some trial and error we determined the value of $T_{flow}=0.5$ for the flow time required in the low temperature calculations.
Since at low temperatures the condensate $\langle\bar\chi \chi\rangle$ as a function of the chemical potential $\mu$ is essentially a step function we show results only for the value $\mu=1.688$ where the condensate abruptly change from the $\mu=0$ value to zero (see \tab{tab:lowTcriticalmu}). On \fig{fig:histograms}  we show a histogram of the 
sampled fields (in tangent space) and the corresponding values of $S_I$.
These histograms clearly shows that regions corresponding to several thimbles are being sampled, despite the high action barriers between them. This was accomplished by taking large values of the proposal, capable to ``jump over" these barriers in a single Metropolis step. As a consequence the acceptance rate is small (around $15\%$). Compounded with the fact that $e^{-i S_I}$ still oscillates significantly, this small acceptance rate leads to a serious loss of efficiency. In the toy  problem discussed in this paper this lack of efficiency can be overcome by brute force (the $N=32$ calculation is the result of a 2 million Metropolis steps).  In any case, these results demonstrate the soundness of parametrizing all thimbles by the tangent space of one of them. A promising research avenue is to substitute the simple Metropolis algorithm we used by another algorithm more adapted to multi-modal distributions (for a review of these methods see  \cite{Neal:1996fk}). 

\section{Conclusion}
We pointed out that the contraction algorithm lends itself to calculations over a whole family of sub-manifolds of the complexified field space, all of which are equivalent to the original real manifold where the partition function is defined. At zero flow ($T_{flow}=0$) this sub-manifold is simply the tangent space to one critical point. On the other hand, in the large flow limit ($T_{flow}\rightarrow\infty$) the sub-manifold is the union of all thimbles contributing to the partition function. This observation has a practical consequence: the contraction algorithm defined around one single critical point can be used to sample {\it all} relevant thimbles, bypassing the tremendous difficulty of identifying all critical points, figuring their contribution to the partition function and sampling them with the proper statistical weights. 

As an example we use this approach in a simple fermionic model, previously studied by several authors.  We showed that for a large parameter region that includes the continuum limit the simplest version of this idea, the no-flow ($T_{flow}=0$) algorithm, is sufficient to produce the correct results, something that eluded previous studies (including the present authors).  For other parameters/models, with more severe sign problems, a more sophisticated algorithm, capable of dealing with multimodal distributions will be required.  This observation establishes a connection between the sign problem and the problem of sampling multimodal distributions. Of course, this is also a hard problem and one might wonder whether this observation can have practical use. This question can be addressed only after  a full implementation is investigated  but we stress that the kind of multimodality expected here is much more benign than, for instance, spin glasses. Indeed, at small values of the coupling constant there is a hierarchy of importance of the different thimbles and only a few of them are expected to contribute significantly while the others are suppressed by small values of $e^{-S_R}$. 

Finally, the fact that the flow maps the main tangent space to the union of all relevant thimbles relies on the fact that the main tangent space is in the same homology class as the original domain of integration $\mathbb{R}^N$. This feature is not really necessary. We can slightly modify  the algorithm by substituting the main tangent plane by $\mathbb{R}^N$ itself. Then one is guaranteed to flow to the correct combination of thimbles. The price to pay is that a larger flow will be required, which is the most computationally intensive part of the calculation.

\acknowledgements

A.A. is supported in part by the National Science Foundation CAREER grant PHY-1151648. 
A.A. gratefully acknowledges the hospitality 
of the Physics Department at the University of Maryland where part of this work was 
carried out.
G.B., P.B., G.R and N.C.W.  are supported by U.S. Department of Energy under Contract No. DE-FG02-93ER-40762.

\bibliography{thimbles} 

\end{document}